\documentclass[fleqn,12pt]{wlscirep}
\usepackage{wrapfig, framed, caption}
\usepackage[utf8]{inputenc}
\usepackage{heuristica}
\usepackage{graphicx}
\usepackage{wrapfig, framed, caption}
\usepackage{tcolorbox}
\usepackage{xr}
\usepackage{hyperref}
\usepackage{xcolor}

\makeatletter
\newcommand*{\addFileDependency}[1]{
  \typeout{(#1)}
  \@addtofilelist{#1}
  \IfFileExists{#1}{}{\typeout{No file #1.}}
}
\makeatother

\usepackage{setspace}
\usepackage[normalem]{ulem}
\useunder{\uline}{\ulined}{}
\usepackage{xparse}
\newsavebox{\fminipagebox}
\NewDocumentEnvironment{fminipage}{m O{\fboxsep}}
 {\par\kern#2\noindent\begin{lrbox}{\fminipagebox}
  \begin{minipage}{#1}\ignorespaces
 \end{minipage}\end{lrbox}%
  \makebox[#1]{%
    \kern\dimexpr-\fboxsep-\fboxrule\relax
    \fbox{\usebox{\fminipagebox}}%
    \kern\dimexpr-\fboxsep-\fboxrule\relax
  }\par\kern#2
 }

\title{Asimov's Foundation -- turning a data story into an NFT artwork}

\author[*1,2,3]{Mil\'an Janosov}
\author[4]{Fl\'ora Borsi}

\affil[1]{Department of Network and Data Science, Central European University, Budapest, 1051, Hungary}

\affil[2]{Datapolis Inc, Budapest, 1112, Hungary}

\affil[3]{Milan Janosov \href{https://linktr.ee/janosov}{https://linktr.ee/janosov}}

\affil[4]{Flora Borsi \href{https://linktr.ee/floraborsi}{https://linktr.ee/floraborsi}}

\affil[*]{janosovm@gmail.com}

\begin{document}

\maketitle


\section*{Abstract}
{\small

In this piece, we overview Isaac Asimov's most iconic work, the Foundation series, with two primary goals: to provide quantitative insights about the novels and bridge data science with digital art. First, we rely on data science and text processing tools to describe certain properties of Asimov's career and the novels, focusing on the different worlds in Asimov's universe. Then we transform the books' texts into a network centered around Asimov's planets and their semantic context. Finally, we introduce the world of crypto art and non-fungible tokens (NFTs)~\cite{nft} by transforming the visualized network into a high-end digital piece of art minted as an NFT. Additionally, to pay tribute to Asimov's devotion to robotics and artificial intelligence, we use OpenAI's Generative Pre-trained Transformer 3 (GPT-3)~\cite{openai1,openai2} to draft several paragraphs of this paper.

}

\vspace{0.5cm}
{\small {\bf Keywords}: text analysis, network science, digital art, NFT, OpenAI GPT-3}

\section{Introduction}

Isaac Asimov's Foundation series is widely considered to be the best science fiction novel series ever written. The first books are centered on the mathematician Hari Seldon, who discovers a way to predict the future, called psychohistory. With the help of this new scientific field, Seldon founded the Foundations – two groups of scientists and engineers – whose purpose was to preserve and maintain human civilization across the galaxy and its countless inhabited worlds.

Earlier research quantified different dimensions of the book industry. These projects covered topics such as understanding the broader evolution of culture, uncovering the hidden formula behind bestsellers, and capturing the key factors of individual career success ~\cite{books, cult, luck}. To highlight the significance of Foundation, here we analyze the career of Asimov, based on the data available on Goodreads~\cite{goodreads}. We construct a time series representation of his career history which enables us to analyze the evolution of his career success over time.

After pinpointing the overall importance of Foundation in Asimov's career, we aim to understand several quantitative characteristics of the series itself. We study the Foundation series as a set of textual data and discuss elementary statistical aspects, such as the world-frequency distribution and differences between separate books~\cite{worldcl}. Then we focus on sentiment analysis and temporal patterns to curate the emotional arcs of Asimov's worlds (different extrasolar planets of the sci-fi series)~\cite{css,arc}. In the last part of the analysis, we tame the relatedness of Asimov's worlds by constructing a network of worlds connected by semantic similarities - mutually co-mentioned words, providing a network view and descriptive statistics of Asimov's universe.

Finally, we connect scientific data exploration to contemporary digital art by transforming a simple network graph into an artistic product, emphasizing how a network graph can convey quantitative insights and bear artistic value at the same time. As a medium, we use the NFT technology that hosts a digital art movement that gained substantial popularity in 2021~\cite{nft,nft2}. This way, our artwork provides a unique experience of the data, and is minted as an NFT on the fitting platform Foundation.app~\footnote{\href{https://foundation.app}{https://foundation.app}}~\footnote{\href{https://foundation.app/@milanjanosov/~/92747}{https://foundation.app/@milanjanosov/~/92747}}.

In addition, tributing to Asimov's pioneering work on popularizing robotics, we use the beta version of OpenAI's GPT-3 engine's API~\cite{openai1,openai2}, which earlier even wrote an entire newspaper article~\cite{guardian}, to draft several parts of this article.

\section{Asimov's career history}
\label{Sec:asimov}

Isaac Asimov (1920-1992) was a professor of biochemistry and one of the most prolific sci-fi writers of all time. His career spans five decades and covers more than 500 pieces, including novels, short stories, and essays he wrote or edited. While he has not been with us for nearly thirty years, his popularity is higher than ever. According to Goodreads~\cite{goodreads}, there are almost three thousand distinct works associated with him and his legacy rated by more than two million people.

Asimov's career kick-started with his robot stories, such as "Robbie" (1939), the introduction of his famous Three Laws of Robotics, and his first blockbuster (1942), "I, Robot" (1942). As of today, "I, Robot" alone has received more than 300k ratings on Goodreads. Yet, it appears that his robot stories were just setting up the scene to his most exquisite work with over a million Goodreads ratings: the Foundation. This epic series covers seven books - the original trilogy (1951-53) and the four sequels (1982-93), from which the latest piece, Forward the Foundation, was published posthumously in 1993.

\begin{figure}[!hbt]
\centering
\includegraphics[width=1.00\textwidth]{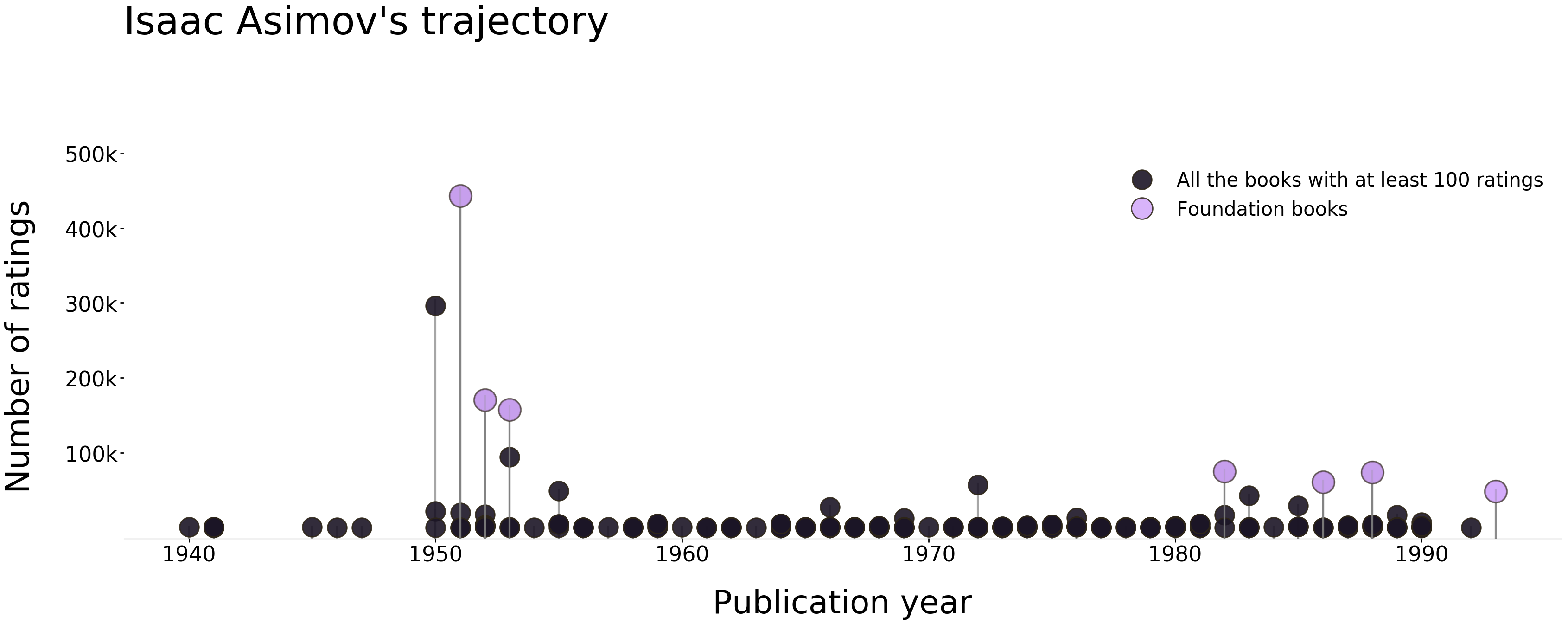}
\caption{Isaac Asimov’s most popular books based on Goodreads ratings. The interactive version of this plot is available here: \href{http://janosov.hu/Fig1.html}{Fig1}.}
\label{fig:fig1}
\end{figure}

When taking a closer look at the evolution of his career based on his Goodreads author profile~\cite{goodreads}, we can construct a data-based time series representation of his career history. In this time series representation, each time event corresponds to the time-stamped publication of a book complemented by the number of ratings it received. The cleaned and filtered career trajectory of Asimov, based on the available data on Goodreads, is shown in Figure \ref{fig:fig1}. Here we only considered books with at least 100 ratings - there were 171 pieces above this threshold. The color coding quickly tells us that indeed, each Foundation book made it to his top 10 most popular books - along with "I, Robot", and two other household names for the sci-fi fans: "The End of Eternity" and "The Gods Themselves".

As outlined by the plot, the last two decades of Asimov's career were more prolific than the first two, while the prominence of the Foundation series is clearly demarcated. The plot also shows that the peak of his career came in 1951 with the release of the first book of the Foundation series, Foundation. Interestingly, the peak of his popularity during the second part of his career, according to Goodreads, coincided with the release of the fourth book of the Foundation series.

To briefly summarize, the plot of the first trilogy of books is centered on the mathematician Hari Seldon, who discovers a scientific field  to predict the future, called psychohistory. With the help of psychohistory, Seldon founded the Foundations – two groups of scientists and engineers – whose purpose is to preserve and expand on humanity’s collective knowledge and to save humanity from a dark age. The plot of the Foundation universe is a story of a galactic empire where a group of humans and robots work together to preserve knowledge and establish a new empire.

\section{Foundation as textual data}

This fall, the long-awaited TV series adaptation of the Foundation series arrives - a sci-fi masterpiece centered by a mathematical theory called psychohistory. This fictional scientific discipline can forecast the future development of large-scale societal systems. Interestingly, today's data and computational social science seem to pursue several similar goals.

When looking at the Foundation through the lens of data science, the question quickly comes: what can we learn about the story as a stream of textual data? To answer this, we combined simple statistical methods, language processing, and network science. As a primary data source, we relied on an open-source digitalized version of the books~\cite{foundation}.

After carefully conducted stemming and lemmatization and removing stopwords (standard text-cleaning steps in natural language processing), we can perform a simple statistical analysis on the vocabulary of the series. For instance, it turned out that there are about 25,000 unique stems mentioned. As a reference, Catcher in the Rye by J.D Salinger has about 4,200 unique words for a word count of 75,000, while Orwell's 1984 uses about 8,600 unique stems in a corpus of 74,000 words. Ranking the most frequently used words, after the verbs 'said' and 'would', the third most frequent one is the name Seldon, while the word 'like' scored at 23th.

Additionally to these numbers, the word "foundation" was mentioned more than 1,600 times, making it the 8th most frequent word in the series; psychohistory with somewhat less than 500 mentions barely makes it to the top 100. There are also three planets, Trantor, Gaia, and Earth, coming up in the top 100 -- three of the about 80 different worlds named~\cite{worlds}. It is also worth mentioning that he introduced the words "robot" and "robotics" in fiction, and often included the term "computer" too. Going to our data set, we can find more than 1000 mentions of these words including their variants. Besides providing interesting insights, these figures also hint that the volume of the series as textual data could suffice for further, more detailed analysis as well.

\section{Asimov's worlds }

\begin{figure}[!hbt]
\centering
\includegraphics[width=1.00\textwidth]{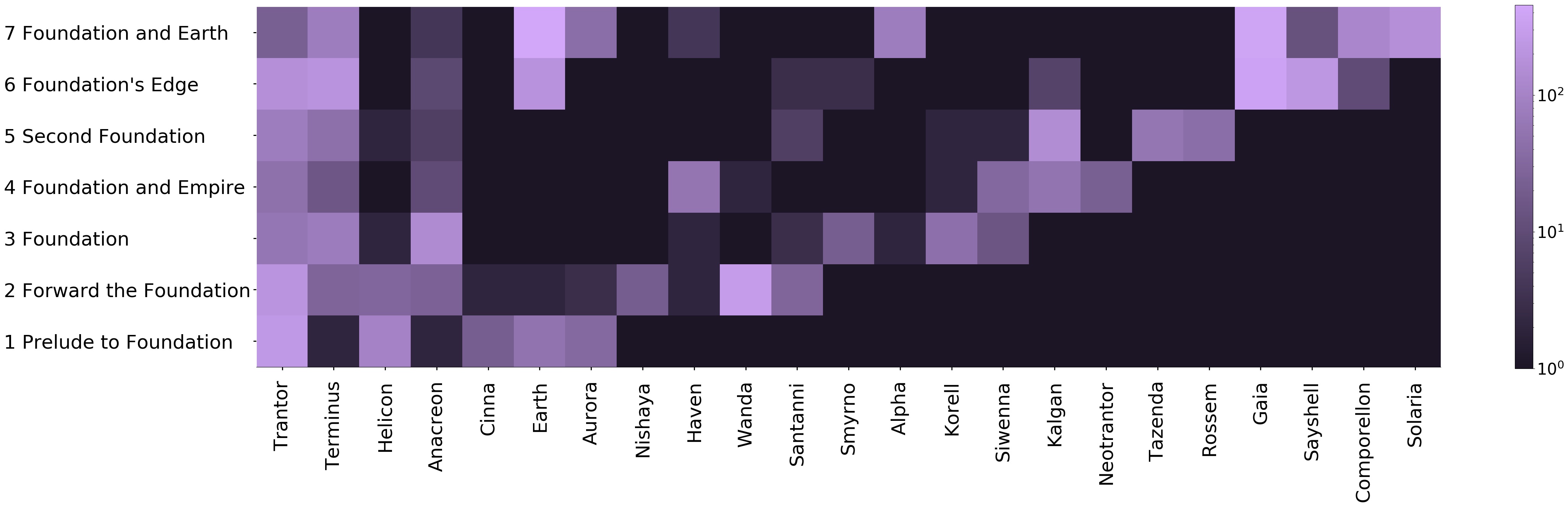}
\caption{Number of named worlds’ mentions within each book.}
\label{fig:fig2}
\end{figure}

The sci-fi series centers around a galactic journey through space and time. In Figure \ref{fig:fig2}, we visualize the number of times each named world was mentioned in the different books - focusing only on those that have occurred at least 20 different times throughout the series. The visualization, color-coding the mentioning count of each world, shows that Trantor, the capital of the First Galactic Empire, and the Terminus, the capital of the First Foundation, have been mentioned in almost every piece. On the other hand, for instance, Helicon, the homeworld of Seldon, is mainly coming up in the first two pieces, consistently to his active years. We can also see that on the Earth and the first colonized extrasolar world, Aurora, the plot is only playing in the early and the later parts of the saga. Additionally, we can observe a diagonal pattern with several planets corresponding storylines typically covering shorter periods. Finally, the Sayshell sector and Gaia within are the most frequently mentioned worlds in the latest episodes - marking where the story of a collective consciousness unfolds.

\begin{figure}[!hbt]
\centering
\includegraphics[width=1.00\textwidth]{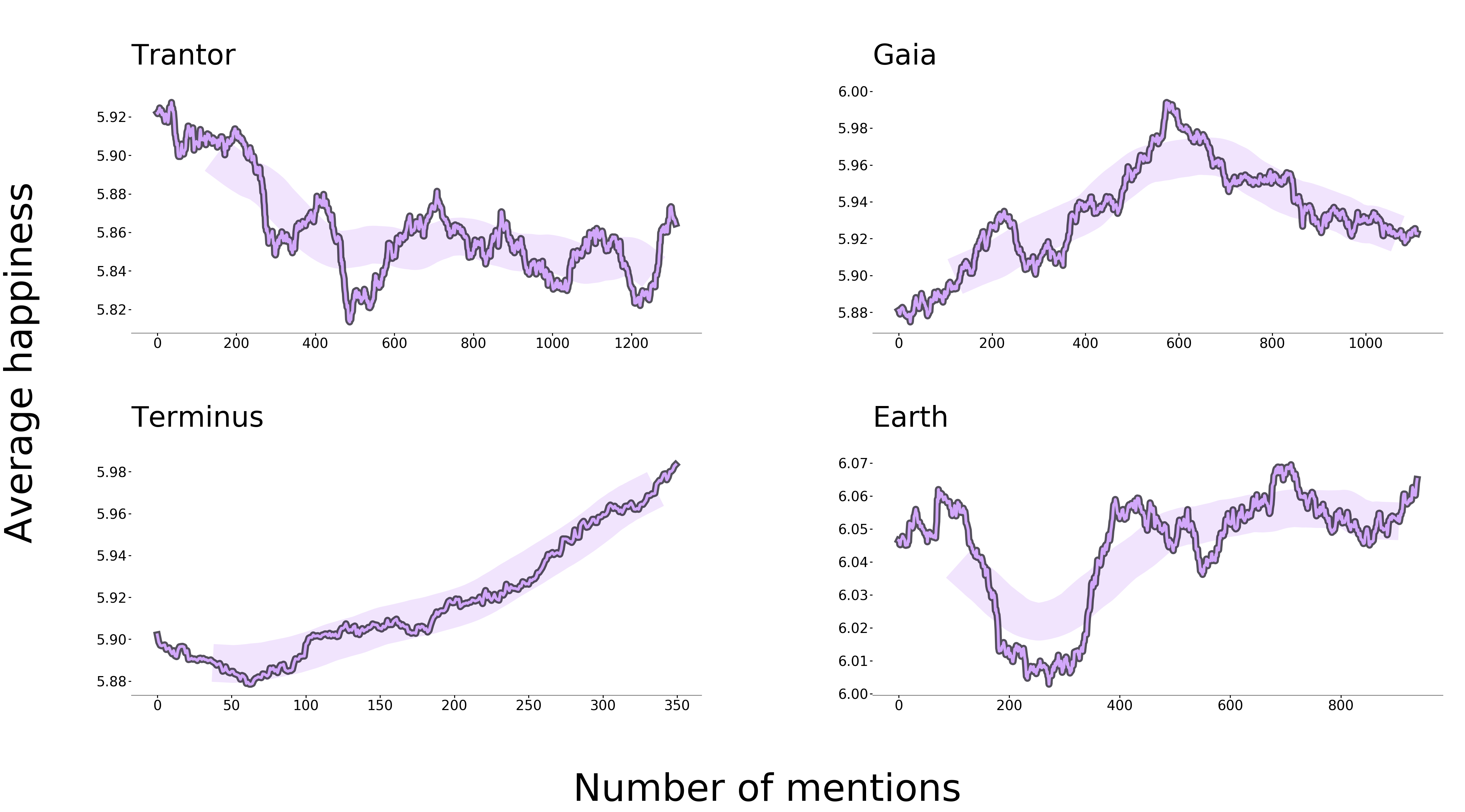}
\caption{The emotional arc of the four planet’s mentions. The lines represent the happiness scores of the mentioning sentences of each world at every mention, while the shaded areas illustrate the binned trends of the arcs.}
\label{fig:fig3}
\end{figure}

If we want to go deeper than mere vocabulary and focus on the entire storyline, there are some further insights we can learn from the text. Earlier research, inspired by Kurt Vonnegut, pointed out that six major types of emotional arcs build up every storyline based on how they unfold over time: {\it i)} rags to riches (rise), {\it ii)} tragedy (fall), {\it iii)} man in a hole (fall-rise), {\it iv)} Icarus (rise-fall), {\it v)} Cinderella (rise-fall-rise), and {\it vi)} Oedipus (fall-rise-fall)~\cite{arc}. To capture these arcs, Reagan et al.~\cite{arc} measured the happiness-level of 10,000-word windows sliding across the text with a sliding steps size of 10 words. They tested their methodology on more than a thousand different books from Project Gutenberg's fiction collection.  Additionally, they measured the happiness level by the aggregation of the individual categorization of each word in the widely-used labMT dataset~\cite{labMT, happiness}. Here we rely on their methodology of extracting emotional arcs by chronologically quantifying the happiness of each world's contexts as the sequence of mentioning sentences.  This methodology then allows us to capture the emotional arc of each world as visualized in Figure \ref{fig:fig3}.

Figure \ref{fig:fig3} visualizing the emotional arcs of four selected worlds tells exciting stories. First, the extended trajectory of Trantor shows a considerable level of fluctuations. However, it is paired with a steady decline -- a typical trend to the genre of tragedy, most commonly exampled by Romeo and Juliet. While the arc of Trantor lines up with the fall of the Galactic Empire, Terminus clearly shows an opposite pattern by its rags to riches storyline.

While these previous two trajectories have one fixed direction, Gaia and Earth show a bit more varying - and interestingly, complementary trends. On the one hand, Gaia follows a rise-fall arc, similar to the Greek myth of Icarus. After analyzing the novel's text, Gaia's turning point seems to be the arrival of the main characters at Gaia and first encountering its collective consciousness and supra organism nature. On the other hand, after initial fluctuations, the happiness path of Earth first falls, then rises (and then saturates) - an arc closest to the one called man in hole, just like The Godfather. The explanation of this arc is less straightforward as the history of Earth covers numerous significant events, such as wars with the colonized worlds, critical overpopulation, and the crust becoming radioactive.

These examples also shed light on how the emotional experience of reading a science fiction book is structured. While the emotional trajectories of different worlds may be more or less the same, the reader is still likely to experience them differently. The reason is that different worlds can activate different projection systems, especially if their stories are short. For instance, the emotional arc of Trantor will be experienced similarly by most people, yet if one of the projection systems is "dark side of the force".

\section{The network of Asimov's worlds}
\label{Sec:network}

\begin{figure}[!hbt]
\centering
\includegraphics[width=1.00\textwidth]{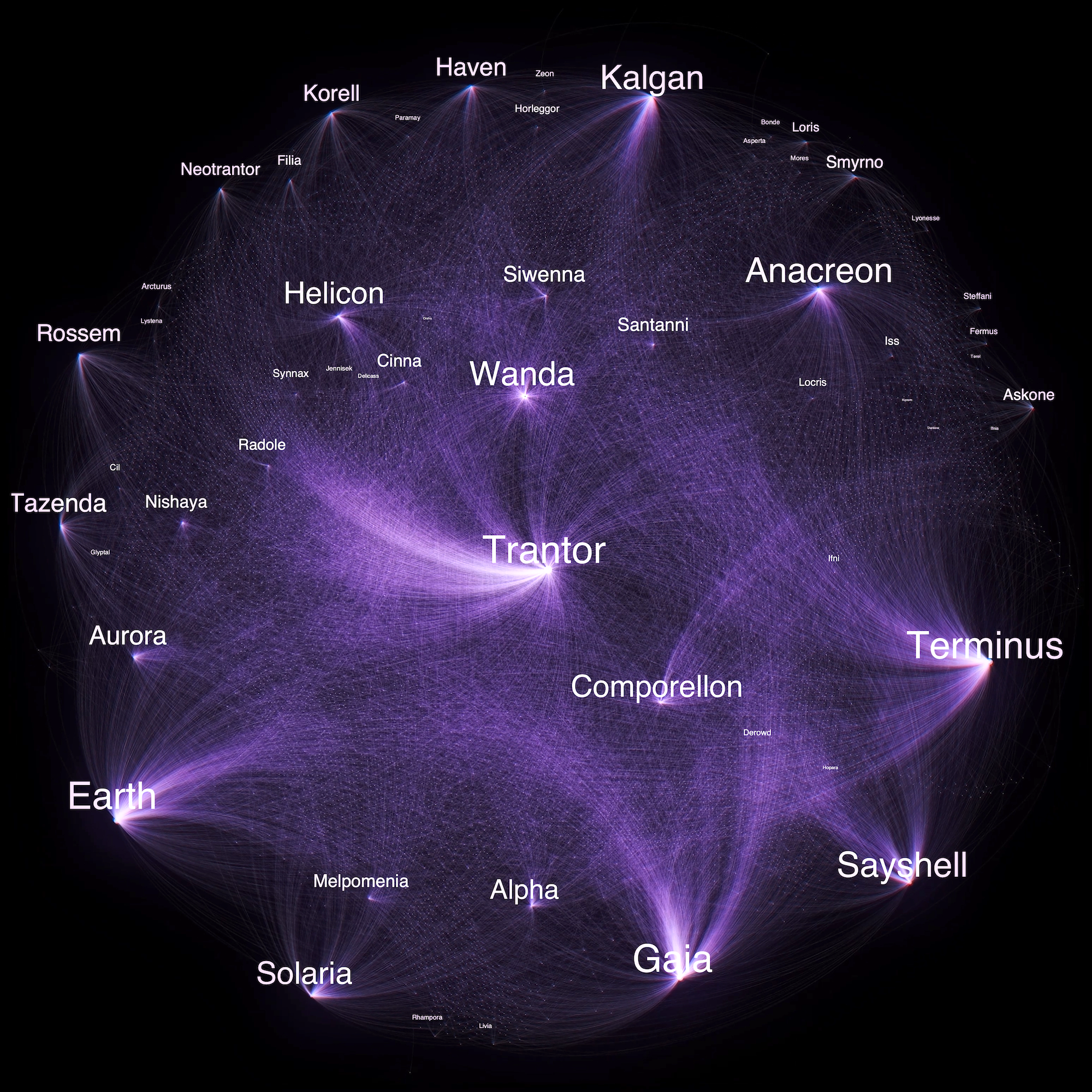}
\caption{The network visualization of Asimov's worlds and co-mentioned words. Each node represents a word or world, where the node size is proportional to its degree count and the edge width measures the number of co-mentions.  The interactive version of this plot is available here: \href{http://janosov.hu/Fig4/index.html}{Fig4}.}
\label{fig:fig4}
\end{figure}

Finally, we transform the universe of the Foundation series into a bipartite network of fictional worlds and words describing them as follows. We consider every word and world remaining after text-cleaning to be nodes, and we connect a planet and a common word if they were co-mentioned in the same sentence. Additionally, the strength of the connection is proportional to the number of co-mentions. Following this definition, we obtain a network of 8,375 nodes (out of which 59 represent planets) and 22,299 links that. The network is visualized in Figure \ref{fig:fig4}.

In this network, the degree distribution shows scale-free characteristics~\cite{scale}, and the nodes with the highest degree coincide with the most frequently mentioned ones: Trantor, Gaia, Terminus, and Earth. The two strongest connections are between the verb 'said' and Gaia, and Earth respectively, while the third strongest link connects the name of Trevize to Gaia, lining up with the themes of the last two books.

The three pairs of most frequently co-mentioned planets are Gaia and Sayshell, Trantor and Terminus, and Earth and Aurora. At first glance,  these statistics seem to be driven by the sheer volume of mentions. To this end, we also uncovered that the correlation between the number of mentions each world has over time (in the sequence of books, shown in Figure \ref{fig:fig2}) and the overall Jaccard similarity of their sets of neighbors is only about $\sim0.18$, meaning that the similar world-profiles of two worlds are not strongly linked with them being mentioned together.

Finally, a community detection algorithm~\cite{arc} further confirms the planet-planet similarities seen in the strongest network links by identifying twelve different network clusters. The network communities reveal that Sayshell and Gaia are parts of the same community, similarly to Trantor and Helicon. In addition, the Earth and its first colony, Aurora, belong to the same network module as well. Somewhat expectedly, and aligning to its nature in the series, Solaria does not cluster with any mentioned worlds. We attached the worlds and their communities in Table \ref{tab:tab1}. Finally, probably the most interesting part about the communities is that they strongly resemble the major themes of the series. For example, the two novel clusters that correspond to the Galactic Empire and the Foundation itself are quite similar to each other.

\begin{table}
\footnotesize
\begin{tabular}{@{}cl@{}}
\toprule
\textbf{Community ID} & \textbf{Member worlds}                                                                                                                                              \\ \midrule
0                     & Terminus, Ifnia                                                                                                                                                     \\
1                     & \begin{tabular}[c]{@{}l@{}}Trantor, Synnax, Santanni, Delicass, Nishaya, Ifni, \\ Hopara, Jennisek, Cinna, Helicon, Derowd, Livia\end{tabular}                      \\
2                     & \begin{tabular}[c]{@{}l@{}}Loris, Glyptal, Korell, Steffani, Horleggor, Locris, \\ Fermus, Paramay, Zeon, Siwenna, Daribow, \\ Konom, Anacreon, Smyrno\end{tabular} \\
3                     & Earth, Aurora, Alpha, Melpomenia                                                                                                                                    \\
4                     & Rhampora, Haven, Lystena, Orsha                                                                                                                                     \\
5                     & Arcturus, Wanda                                                                                                                                                     \\
6                     & Askone                                                                                                                                                              \\
7                     & Comporellon, Iss                                                                                                                                                    \\
8                     & Filia, Neotrantor                                                                                                                                                   \\
9                     & Rossem, Kalgan, Terel, Radole, Mores, Asperta, Lyonesse, Bonde, Tazenda, Cil                                                                                        \\
10                    & Sarip, Voreg, Zoranel, Gaia, Sayshell                                                                                                                               \\
11                    & Solaria                                                                                                                                                             \\ \bottomrule
\end{tabular}
\caption{The network communities and the different worlds that fall into them.}
\label{tab:tab1}
\end{table}

\section{The NFT artwork}

Non-fungible tokens (NFTs)  are chunks of data stored in a digital ledger, in our case, on the Ethereum blockchain. Due to the technological characteristics, every NFT is certified to be unique and interchangeable~\cite{nftp}. This allows NFTs to be traded and collected, i.e., original physical artwork and other collectibles. By today, there is an entire ecosystem of NFTs with various trading platforms.

\begin{figure}[!hbt]
\centering
\includegraphics[width=0.800\textwidth]{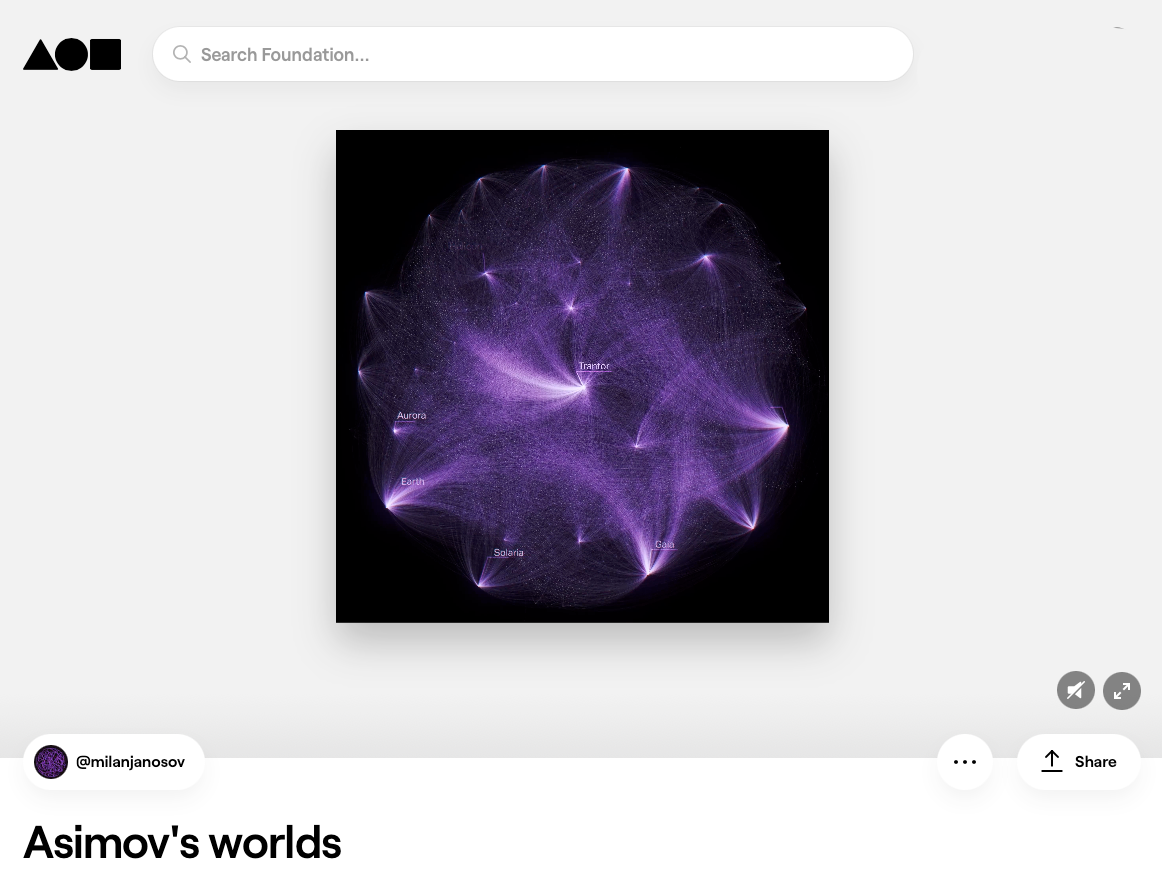}
\caption{Asimov's worlds minted as an NFT artwork at \href{https://foundation.app/@milanjanosov/~/92747}{https://foundation.app/@milanjanosov/~/92747}.}
\label{fig:fig5}
\end{figure}

With this paper, we aim to present a scientific data visualization and simultaneously transform it into a digital piece of art by combining the data visualization with different digital visual effects, motion graphics, and unique audio. We minted the artwork as an NFT on Foundation.app titled as {\it Asimov's worlds} also shown on Figure \ref{fig:fig5}

\section{Summary}

This article aimed to explore the universe of Isaac Asimov's Foundation from a network and data science perspective and produce a data-driven digital piece of art present in the crypto art space. First, we analyzed the career patterns of Asimov to obtain a quantitative framing of his masterpiece. Second, we turned to the texts of the seven books into a data set and conducted a brief statistical analysis on the world-frequency profile of the series. Then, we dove into the characteristics of Asimov's worlds, and in particular, computed the different emotional arcs the key planets have. Finally, we built a network of the co-mentioned worlds and words and transformed the data visualization into a crypto art piece minted as an NFT. Additionally, we used an OpenAI's tool to draft a measurable proportion of this article.

\section{Data accessibility}
Supplementary files associated with this study can be found at
\href{https://github.com/milanjanosov/Foundation}{https://github.com/milanjanosov/Foundation}.

\section{Authors' contributions}
F. B. and M. J. proposed the idea of the study. M. J. performed the data collection and analysis and wrote the manuscript. F. B. created the NFT artwork and composed the soundtrack. OpenAI's GPT-3 engine produced about 10\% of the words in the text.


\section{Acknowledgement}
The authors wish to thank Manran Zhu for introducing the beta version of OpenAI GPT-3 and Ágnes Diós-Tóth for the careful review of the manuscript.

\end{document}